\documentclass[letterpaper, 10pt, conference]{ieeeconf}      

\IEEEoverridecommandlockouts                              

\usepackage{graphics} 
\usepackage{epsfig} 
\usepackage{mathptmx} 
\usepackage{times} 
\usepackage{amsmath,amssymb,amsfonts}

\usepackage[utf8]{inputenc}
\usepackage[T1]{fontenc} 
\usepackage{graphicx}
\usepackage{textcomp}
\usepackage{xcolor}

\usepackage{xspace}
\usepackage{hyperref}
\usepackage{float} 
\usepackage{units}
\usepackage{multirow}
\usepackage{hhline}
\usepackage{array}
\usepackage{color, colortbl}

\usepackage{tikz}
\usetikzlibrary{shapes.geometric}
\usetikzlibrary{calc}
\usetikzlibrary{positioning}
\usetikzlibrary{intersections}

\usepackage{pgfplots}
\usepgfplotslibrary{fillbetween}

\usepackage[section]{algorithm}
\usepackage{algpseudocode}
\usepackage{xspace}
\usepackage{hyperref}

\usepackage{tabularx,ragged2e,booktabs,caption}
\newcolumntype{C}{>{\Centering\arraybackslash}X} %

\title{ \bf Virtual Reality Assisted Human Perception in ADAS Development:\\ a Munich 3D Model Study}

\author{Felix Bognar$^{1,2}$, Markus Oster$^{2}$, Herman Van der Auweraer$^{1}$, Tong Duy Son$^{1}$ \\
\vspace{0.1cm}\\
{$^{1}$ Siemens Digital Industries Software,  3001 Leuven, Belgium}  \\
{$^{2}$ University of Applied Sciences Munich, 80333 München, Germany}
\thanks{This work was conducted within the thesis internship of Felix Bognar at Siemens Digital Industries under supervision of Dr. Son Tong, and in collaboration between Siemens Digital Industries Software and University of Applied Sciences Munich. The work has been finalized with the thesis of Felix Bognar supervised by Prof. Dr. Markus Oster at the Faculty of Geoinformation. Email: {\tt\footnotesize son.tong@siemens.com, felix.bognar@t-online.de, markus.oster@hm.edu}.}
}

\begin{document}
\maketitle
\thispagestyle{empty}
\pagestyle{empty}


\begin{abstract}

As the development of autonomous driving (AD) and advanced driver assistance systems (ADAS) progresses, the relevance of the comfort of users is gaining increasing interest. It becomes significant to test and validate perceived comfort performance from the early phase of system development before driving on roads. Most of the present ADAS test procedures are not efficient in performing such comfort evaluation. One of the main challenges is to integrate high-quality, realistic and predictable virtual traffic scenarios into an ADAS testing framework that has physics-based sensors capable of sensing the virtual environment. 

In this paper, we present our development of a virtual reality based ADAS testing framework that enhances human perception evaluation. The main contribution relies on three aspects. First, we introduce our development of a large and high-quality (realism, structure, texture) 3D traffic model of the Munich city in Germany. Second, we optimize the 3D model for virtual reality purpose, and real-time capable for human-in-the-loop ADAS testing. Finally, the model is then integrated into an ADAS framework for testing and validating ADAS functionalities and perceived comfort performance. The developed framework components are presented with illustrative examples.

\end{abstract}

\section{Introduction}

In the development of autonomous driving (AD) or advanced driver assistance systems (ADAS), validating the safety and comfort perception by the users of these systems represents an important role. This requires extensive testing efforts during the development process. These tests can be conducted in several ways, either with real physical vehicles on closed-off terrain or in simulated environments. The ultimate objective is to achieve optimal performance in terms of safety, comfort, and energy consumption \cite{TONG19, sontra2017}. 

Physical testing is essential to demonstrate the safety of autonomous driving and build trust and acceptance with end-users in the final step \cite{TUV21}. However, particularly safety-relevant scenarios can only be realistically performed with real cars in limited cases due to the high risk of accidents and effective use of resources. Hence, ADAS testing is often performed in a virtual, simulated 3D environment throughout development stages. These simulation software tools offer the possibility to generate realistic 3D traffic scenarios with static and dynamic traffic objects. In addition, all sensors required for the ADAS algorithms, such as cameras, radar, or LIDAR sensors, which are envisioned on the test vehicle, can be virtualized and simulated. Some simulation tools can also render real-time 3D visualizations of the test scenarios, that enable to produce photo-realistic visualizations using technologies such as 3D modeling, physics-based rendering and materials, ray tracing, and dynamic lighting \cite{Prescan, Carla}. 

However, both real and simulation based testing methods run into limitations when validating human perceived safety and comfort perception. Physical closed-off terrain tests are often relatively static and lack of complexity compared to real traffic, while simulation often does not provide a sufficient realistic environment. In real tests, the user can feel the driving dynamics of the test vehicle, but he/she sees a rather unnatural scenario with crash test dummies on an empty closed-off area. In addition, the test subject would be exposed to enormous risks in the event of technology failure in such a test procedure. Simulations, on the other hand, can be very complex in design that contain dynamic elements without risking an accident. The user's perception of comfort requires realistic traffic and driving dynamic models, which play a decisive role in simulation-based testing performance. 

\begin{figure*} 
    \centering
    \includegraphics[width=14cm,height=15.5cm]{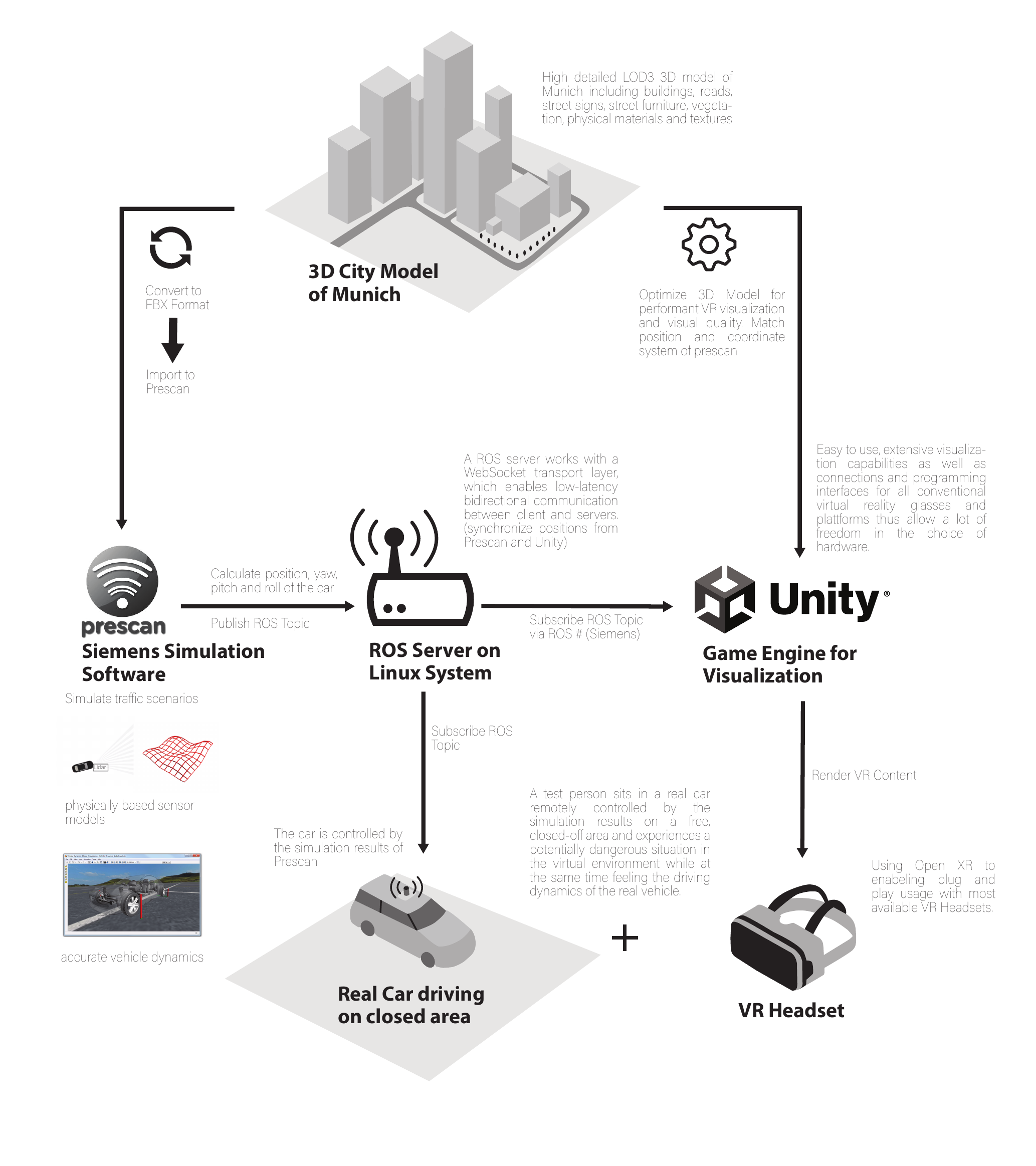}
    \caption{Framework illustration of the modular communication between the simulation (Prescan) for physics-based sensors and ADAS algorithms development on the left and the virtual reality (VR) visualization based on a game engine (Unity) on the right via a ROS server. The high-quality 3D Munich city model is imported to both environments.}
    \label{fig:DetailedToolchain}
\end{figure*}

In this work, we attempt to improve human perception performance in ADAS/AV testing via developing and combining high quality traffic and vehicle dynamics models. The framework is capable for both physical and virtual based testing. At first, the framework relies on the standard Model-in-the-Loop (MiL), Hardware-in-the-Loop (HiL) and Vehicle-in-the-Loop (ViL) testing architectures in the automotive industry. The vehicle is controlled either on a driving simulator or in a closed-off area under a simulated traffic environment. In addition, we provide additional layers aiming at qualitative human perception aspects, including:
\begin{itemize}
  \item a large and high-quality 3D traffic model, with an example of the Munich city, Germany;
  \item 3D model development in a game engine for virtual reality (VR);
  \item coupling the 3D model with ADAS simulation software including physics-based sensor models (lidar, camera), high-fidelity vehicle dynamics, and ADAS algorithms (perception, planning, control).
\end{itemize}
To this end, we aim to co-simulate a large 3D Munich model of realistic buildings, roads, traffic signs, ... with high-quality materials and textures into the ADAS simulation toolchain. Since the model is developed and calibrated from a real-life environment in highly significant details, it enhances the credibility of virtual testing. Virtual reality visualization can be implemented with the simulation software itself or with a game engine, whereby game engines usually allow more configuration of the display quality and performance. In addition, the implementation using a game engine represents a platform-independent solution and enables VR visualizations even with simulation software that does not demonstrate such functionality. For synchronization of the simulation with a game engine, the interface can also be used via local network. The platform architecture is shown in Fig. \ref{fig:DetailedToolchain}.

\section{3D Munich City Model}

As part of an elective subject in the field of geoinformation at the Munich University of Applied Sciences, a 3D model of the Munich district of Maxvorstadt had been developed since 2010 at the Faculty of Geoinformation (course of studies: GeoVisual Design - Geomatics). The 3D model and it's building workflows are constantly being expanded and improved since 2010, and have been conceptualized by Prof. Dr. Markus Oster as part of the course. The model includes buildings, streets, street furniture, vegetation, parks and other details of the corresponding area. In particular, the entire model contains textures and materials based on the real world, which are not simply mapped with texture maps over the entire surface, but are assigned to the individual elements as material. This significantly increases the quality of the visualization, especially for perceiving the simulated environment from a car driver or pedestrian perspective.  This is common in the game engine, where the materials should allow a high display quality and thus realism even at a short distance from the virtual camera and not pixelate. In the usual automated UV mapping of entire facade sides on 3D bodies, this high level of quality can only be achieved with bird's eye views or at a greater distance. In addition to the display quality, the material definition of individual elements also offers the advantage of being reusable in the city model and, when necessary, being able to adapt materials with only simple modifications. For example, in the case of a roughcast of a house facade, the material model can be adapted easily via relief maps in terms of roughness and/or via color changes.

\begin{figure}

\vspace{0.25cm}

\centering
    \includegraphics[width = 0.47\textwidth]{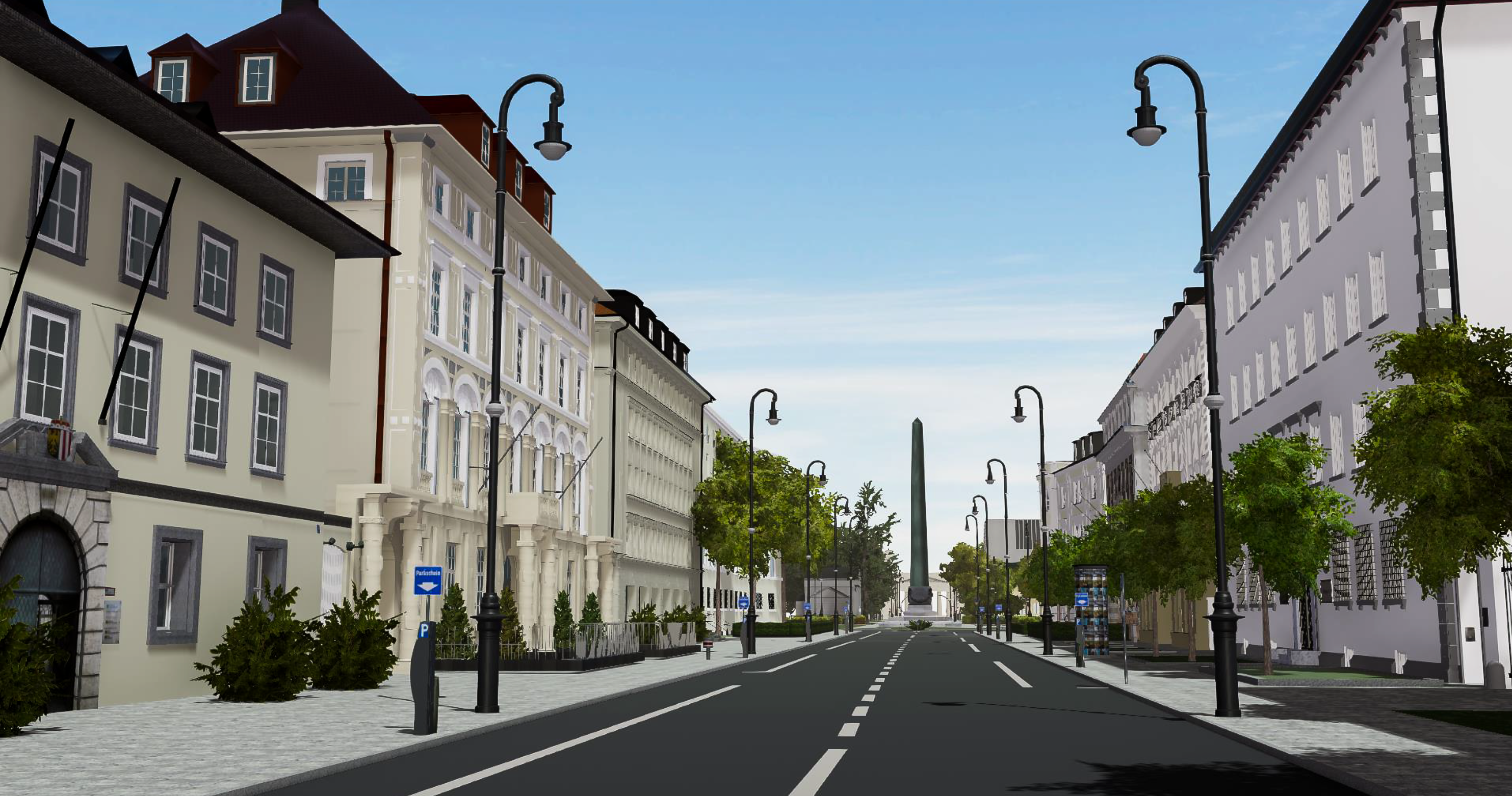}

\vspace{0.2cm}

    \includegraphics[width =  0.47\textwidth]{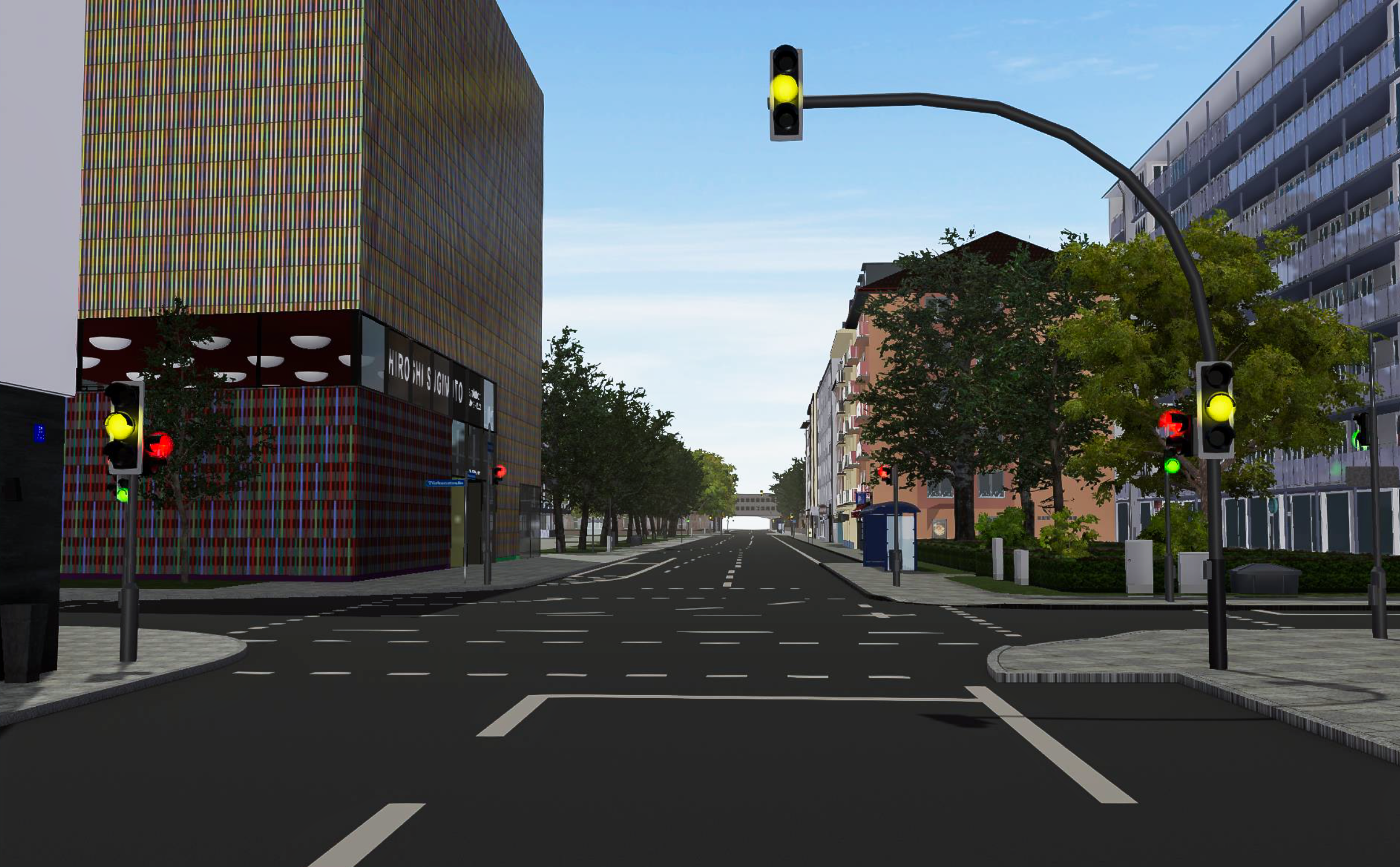}

\vspace{0.2cm}

    \includegraphics[width= 0.47\textwidth]{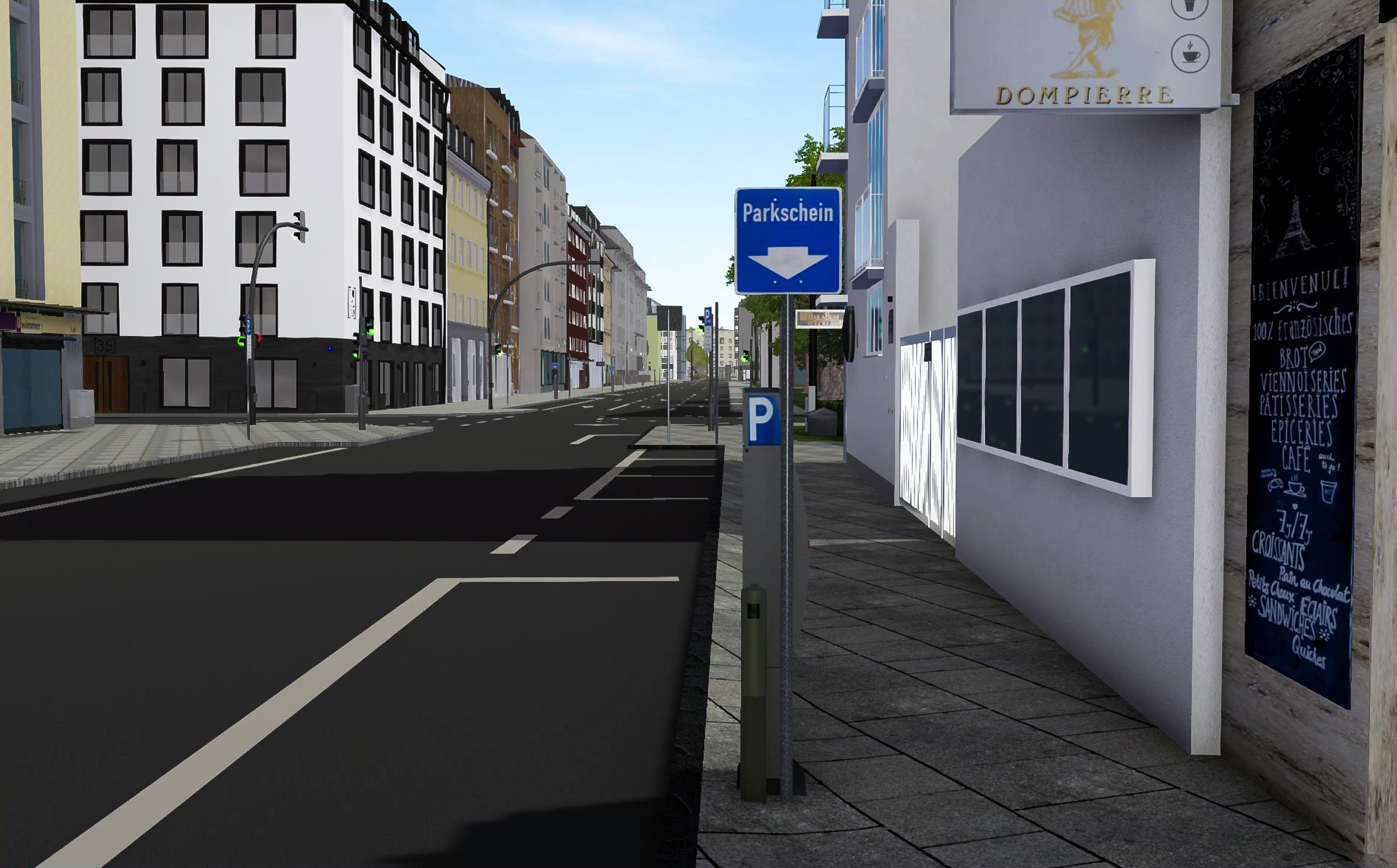}
    \caption{Developed 3D Munich model: Brienner Str. , Theresienstraße, Türkenstraße.}
    \label{fig:MunichModelView}
\end{figure}

\begin{figure*} 
    \centering
    \includegraphics[width=1.0\textwidth]{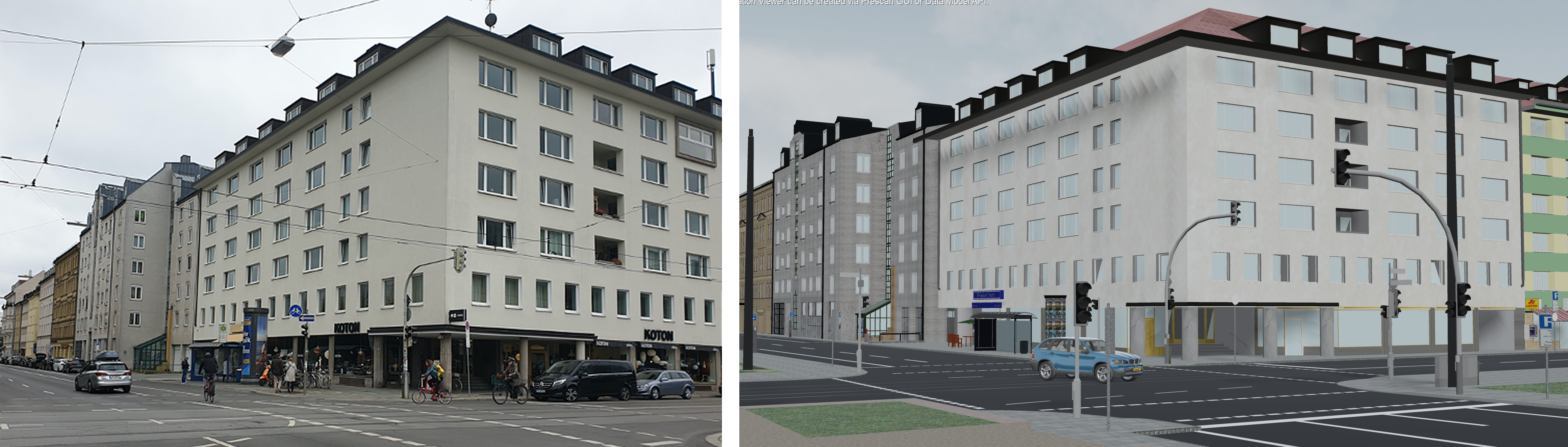}
    \caption{Real photograph vs rendered image (imported to Prescan) of a building in Gabelsbergerstraße Street.}
    \label{fig:MunichModelComp}
\end{figure*}

The 3D models of the buildings can be referred to as Level of Detail 3 (LOD3) models, which cover maximum building details visible from the outside \footnote{not to be confused with the LOD designation in Unity, where LOD0 represents the highest level of complexity.}. In Fig. \ref{fig:MunichModelView}, some examples of the modeled Munich streets are shown, demonstrating the details and quality materials, textures. In addition, Fig. \ref {fig:MunichModelComp} compares the model with a real life photo. 

The 3D model has been created as an interactive city model in the game engine Unity that additionally allows access to different semantic data. LOD2 data of the State Office for Digitization, Broadband and Surveying in Munich (LDBV) \cite{LDBV} has been interactively integrated into the city model where the individual developed buildings are deposited to the building-specific data of LDBV. Buildings which are not yet modeled can be filled using simple LOD2 models to obtain a closed cityscape. Navigation within the city model is implemented using a first-person controller with free movement within the virtual environment. Furthermore, there is a menu control to change resolution, render quality and other settings. We can adapt the settings depending on hardware configuration for proper visualization.

Most importantly, the individual 3D models that make up the city model have each been developed and created by various students using the modeling software 3ds Max \cite{3dsMax} or Blender \cite{Blender}. The basis of the modeling is so-called facade photos, photographs of the respective building, taken using an single-lens reflex (SLR) camera and tilt-shift lenses. We use the tilt-shift lens to reduce the perspective distortions of the photographs, which can occur due to the low shooting angles in relation to the height of the building. In addition, reference dimensions of the building are measured, where the nearly distortion-free images can be used as a measurement basis in the modeling process. For this purpose, the pixel values of the image are set in relation to the reference dimensions. This ratio factor is then used to calculate further dimensions from the image.

In general, one photograph is taken for each facade side of the building to derive the dimensions. Moreover, separate images of specific building details are taken to increase the level of complexity. This improves the overall accuracy of the dimensions, as additional control values are available to validate the result. In some cases, the dimensions of parts of the building that cannot be seen or measured are calculated using either remote sensing data or building plans. Particular detailed parts of buildings are also created using a photogrammetric method. This method, also called photo-scanning, is based on taking several images from different points of view. Depending on the object characteristics, this procedure requires 20 to 250 images of the corresponding object. 

In the next step, the individual images are computed into a highly detailed point cloud object and consequently a 3D mesh. The mesh of the 3D model must then be optimized to reduce the number of polygons in areas of the model where they are not needed - but still keeping the same high fidelity quality. This results in a considerable performance advantage, especially with an extremely large number of 3D models. These photogrammetrically derived detailed elements, such as statues or decorative objects, are then integrated into the manually constructed models of the buildings in the correct position. In the following subsections, we will present our developed 3D Munich model quality in detail. We will also compare with other conventional modeling methods concerning structure and texture qualities.

\subsection{High structure quality} 

\begin{figure}[b]
    \centering
    \includegraphics[width =  0.485\textwidth, height =  3.80cm]{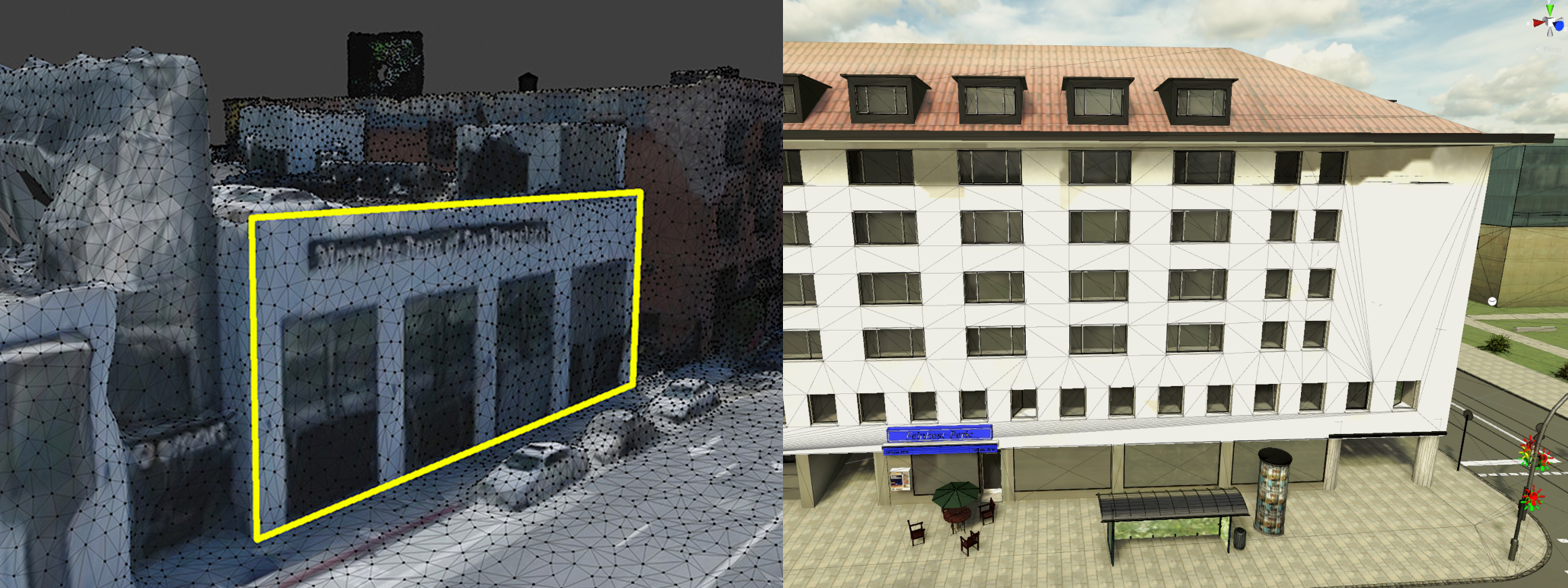}
    \caption{Example comparison: from laserscan \cite{3DScann20}} vs our 3D model.
    \label{fig:TextureQuality}
\end{figure}

Conventional automated processes like photo or laser scanning create a grid of vertex points which the object is covered in. The density of the grid, and respectively the amount of the vertex points, determines the resolution. These grids can be algorithmically filtered to reduce the number of vertex points. However, the absolute number of vertex points is always directly related to the quality and level of complexity of the 3D model. This leads to much larger file sizes and possible errors in the creation  process (see an example of laserscan model in Fig. \ref{fig:TextureQuality}).

In our development, the number of vertex points and resulting polygons of the 3D models are greatly reduced compared to the fully automated methods thanks to manual creation or revision. We focus on the relationship between the number of vertex points and the level of complexity of the real building. The density of vertex points varies adaptively depending on the complexity of the object or within parts of the object (see an example or our model and comparison with laser scanning method in Fig. \ref{fig:TextureQuality}).

\subsection{High texture quality}
The authenticity of simulation environments is always a critical question in visualization and human perception. This lack of authenticity is performed as the texture of the rendered image looks not realistic as the real world \cite{LouDeng}.
Automated processes for creating 3D environments for ADAS simulations often rely on photo textures. This is mainly due to the structural differences in the resulting 3D models. 
For example, in laser scanning, the surface of an object is captured very precisely and this 3D shell covers the object structure. However, this highly accurate 3D shell has no information about the texture, material, and especially the thickness of the object. Since the thickness of materials is a determinant influence on their physical properties, it complicates or excludes the use of physically correct materials for 3D models or meshes with these properties. Consequently, the 3D shell is usually overlaid with simple photos in simulation applications which can be visually appealing. This method freezes a situation of the real world and all associated environmental parameters (sun position and intensity, weather,...) permanently, which is why only these exact environmental parameters need to be used in later applications. Additionally, these photos contain shadows, reflections, and other artifacts which are not consistent with each other and or with the real world. This limits considerably authenticity and comparability with the real world.
 
In our manual development process of the Munich models, an object is not modeled as a 3D shell but as a full geometric body. This means that the width, height, and depth of each object are defined and modeled. It enables physical materials to be assigned in each part of the 3D model referring to the corresponding real-world surfaces. Physical materials, furthermore, enable the calculation of material-specific properties such as reflection and refraction for the specific environmental parameters specified in the simulation. This is even possible in real time using ray tracing. Therefore, such 3D models can be used with dynamic environment parameters.

\begin{figure*}
    \centering
    \includegraphics[width =  1.0\textwidth]{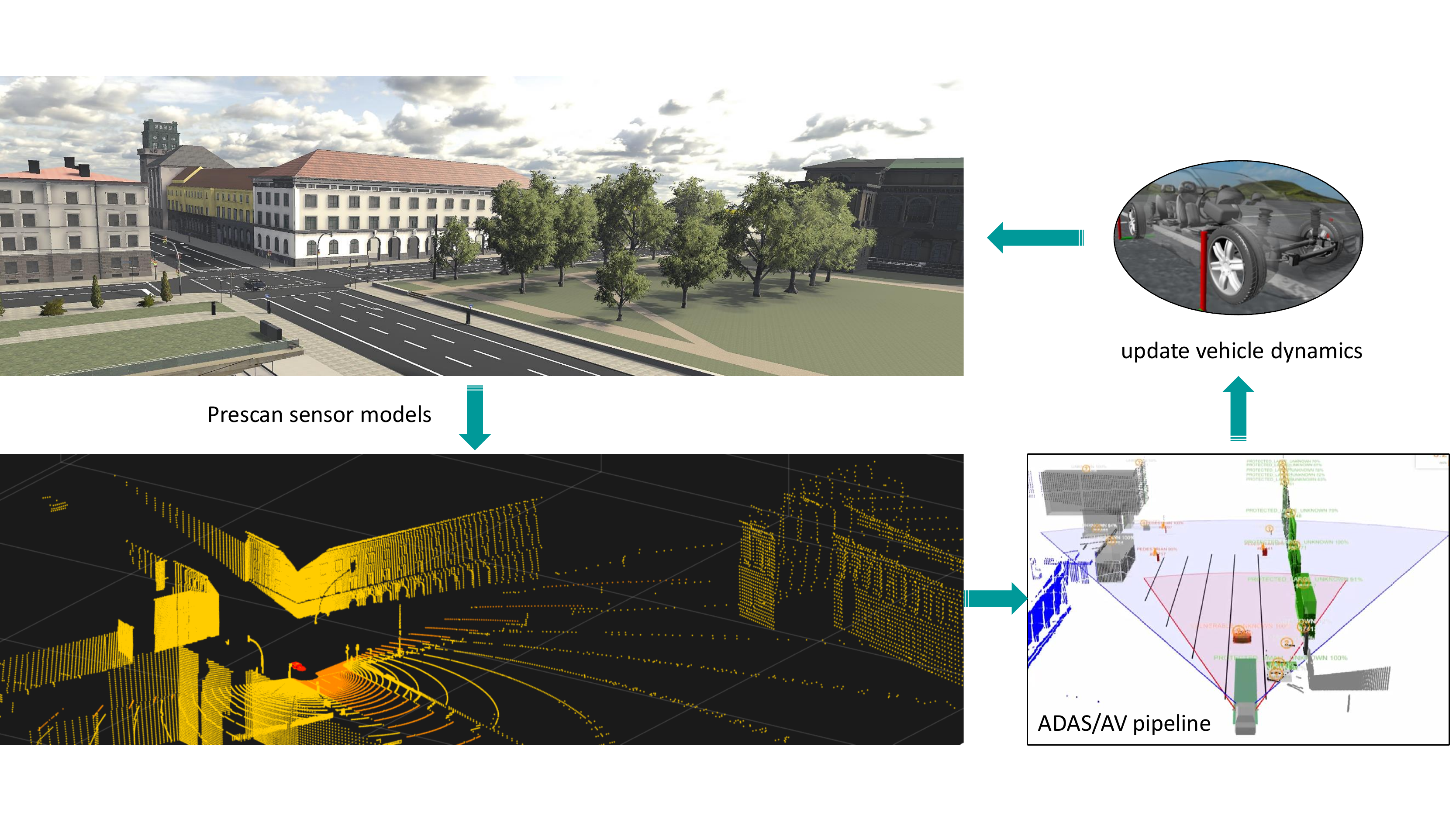}
    \caption{Prescan lidar point clouds on the 3D Munich model, which are fed into an AV stack (perception, planning, control), and the vehicle dynamics driving on the 3D model.}
    \label{fig:PrescanClosedloop}
\end{figure*}

Our solution and efforts might be time-consuming, but lead to photorealistic simulations and achieve authenticity and comparability to the real world environments (see again Fig. \ref{fig:MunichModelView},  and Fig. \ref{fig:MunichModelComp}).

\section{ADAS Testing using 3D Munich Model}

Given the developed 3D Munich model, our next step is integrating the model into an ADAS testing framework, aiming for testing comfort perception of users. The framework consists of three main parts: the Munich model, virtual reality visualization for user, and the ADAS pipeline connecting with the Munich model. The ADAS system includes high-fidelity vehicle dynamics,  physics-based sensor models, and the algorithms to sense, plan and control the vehicle in the traffic environment. Fig. \ref{fig:PrescanClosedloop} represents an example of the complete MiL closed loop testing with vehicle dynamics.

As an example for the testing activity, a person can sit in a physical car controlled by the ADAS system on a free, closed-off area like a proving ground track, and views the virtual traffic environment by means of a virtual reality glass. The subject thus experiences various traffic scenarios in the virtual environment and at the same time experiences the driving dynamics of the real vehicle. Instead of a real car, once may use a driving simulator. The key differentiator here is a high-quality and realistic virtual environment, depicting very close to real streets in the Munich city during testing. This provides significant benefits as the testing environment is familiar and expectable to the users. 

\subsection{ADAS simulation pipeline}\label{AA}

\begin{figure}

\vspace{0.25cm}

    \centering
    \includegraphics[width =  0.47\textwidth]{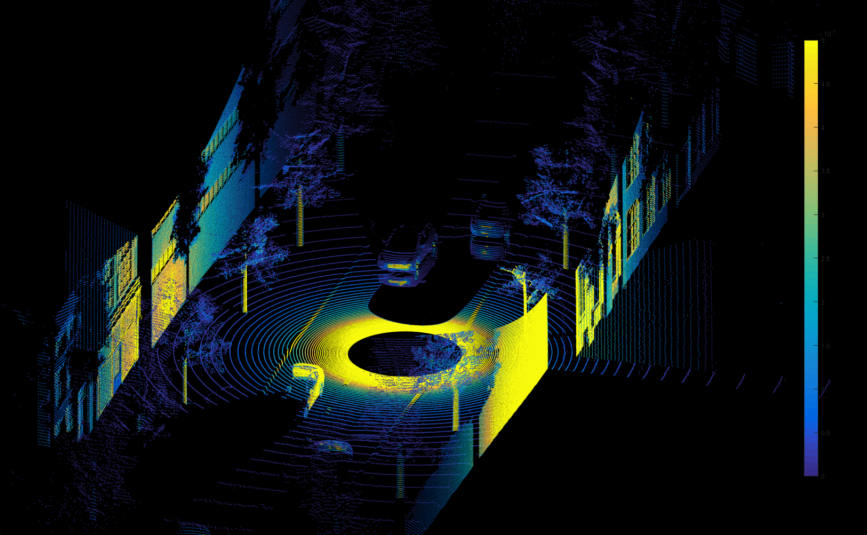}
    \caption{Prescan physics-based lidar sensor simulation}
    \label{fig:PrescanPhysicsLidar}
\end{figure}

The 3D Munich model is imported to the Siemens traffic simulator software Simcenter Prescan. Prescan provides a simulation environment for the development of ADAS and autonomous driving vehicles at different stages. It enables automotive players to test their intelligent systems in a variety of traffic conditions manually or automatically via automatic scenarios generation. In addition, Prescan allows users to develop their own sensors, controllers, and collision warning functions \cite{KAUR21}. The test vehicle can be equipped with different physics-based sensors (lidar, camera, radar, ...). Fig. \ref{fig:PrescanPhysicsLidar} demonstrates the physics-based lidar sensor model. The data streams from these virtual sensors serve as input to the algorithm. In addition, high-fidelity and accurate vehicle dynamic characteristics can be integrated into the simulation. The vehicle model has a complete chassis including powertrain, braking, suspension, steering components producing realistic motion perception for users during the VR testing.

Prescan has been used to validate automotive XiL development cycle: MiL with high-fidelity vehicle dynamics, HiL with vehicle actuation and embedded platform, and vehicle testing (ViL). The autonomous driving environment contains both simulation and physical proving ground environments, see \cite{bruggner2021, flavia2021, allamaa2022}. 

%
%

\subsection{Middleware}

To bridge between different components in the developing ADAS framework, in particular between the ADAS pipeline and the virtual reality visualization, we use Robot Operating System (ROS) as an open-source operating system \cite{ROS21A}. ROS provides a collection of software and programming modules that can be used to create robot applications. The available services are similar to an operating system and provide hardware abstraction, device drivers, utility functions, inter-process communication, and packet management. Three types of communication are available and used in this work, including synchronous communication via ROS services, communication via data stores on a parameter server, and asynchronous streaming via topics.

Communication between the components is implemented by means of a Rosbridge server on a Linux system over local network. The ROS server works with a WebSocket transport layer, which enables low-latency bidirectional communication between clients and servers. This allows sending position data from the simulation to the vehicle in near real-time. Various approaches are conceivable for the implementation of close to real-time virtual reality visualization based on data exchange via local network.

 \section{Virtual Reality Assisted Human Perception}

Enhancement of the visualization capabilities is an essential requirements to obtain high quality human perception. Most simulation tools have the possibility to render real-time 3D visualizations of the test scenario. However, they often offer only limited configuration options and no native support for virtual reality (VR) glasses. To enable high-quality VR visualization, these needed to be optimized manually by engineers. Such process and its feasibility depends on the respective software and is in most cases not directly transferable to other simulation software. This is a platform-dependent solution. Moreover, there could also communication delays between the simulation and the vehicle - real or driving simulator - due to the transmission rate of the local network. Assuming that the simulation combined with the VR visualization executes in real time and the position data is sent to the real vehicle via a network, the speed of the data interface would always be visible as a delay between visualization and the movement of the real vehicle. This could lead to irritation, loss of immersion or sickness of the participants.

\subsection{External visualization through game engine}

In order to generate the visualization, especially in the VR context, independently of the respective simulation software, using a game engine as an external solution is an option. Here, the VR visualization is rendered completely by the game engine on a separate hardware from the simulation computing hardware. The vehicle states (positions, velocities,...), that are provided by the simulation software on the ROS server, can be exploited in this case to synchronize the simulation with the visualization of the game engine. In order to realize such settings, the basis of such a setup is a detailed 3D model, which must be available in both the simulation software and the game engine.

The positioning of objects in the simulation software as well as in the game engine is based on a coordinate system. This makes it possible to place and synchronize the 3D model and the dynamic objects in both software products at the same, unique position. Game engines are software frameworks that are primarily used to develop video games. In recent years, the video game industry has become a huge market with large amounts of investment are being spent on the development of game engines. In contrast to their development costs, the price of the final product is very low compared to a professional 3D visualization/animation program \cite{FRIESE08}.

The separation of simulation from visualization on the hardware level, which is associated with the use of game engines for combined algorithm tests, also brings some performance advantages. Visualizations in virtual reality are very computationally intensive, especially at high-quality and resolution. Parallel computation of the simulation and the VR visualization could lead to performance improvement. 
In addition, delays in data transferring between the software modules are compensated or reduced, since the position data of the simulation is shared on a Linux server and can be read out simultaneously by the game engine and the real vehicle (see Fig. \ref{fig:DetailedToolchain}). If the visualization is calculated on the same hardware as the simulation, there could be delays between visualization and the movement of the real car due to the limited transfer rate of the network.

\subsection{Game engine}

When choosing the game engine, Unreal Engine or Unity comes into question. They are the most widely used available game engines with high-performance and easy-to-use visualization interfaces. 
We selected Unity because a high-quality 3D model is required for the development of the test procedure; in particular, University of Applied Sciences Munich has been working on the creation of a detailed 3D image of the Maxvorstadt district in Munich using Unity since 2010. As already mentioned, the 3D city model represents the data basis in Prescan as well as in the Unity virtual reality scene. For this purpose, a copy of the current project has been created in the first step. In addition to the city data, a 3D model of a vehicle has been integrated into the Unity project, which represents the test vehicle and to be synchronized with the test vehicle in Prescan using vehicle states data (see Fig. \ref{fig:PrescanClosedloop}).

\subsection{VR integration framework}

\begin{figure}

\vspace{0.25cm}

    \centering
    \includegraphics[width = 0.47\textwidth]{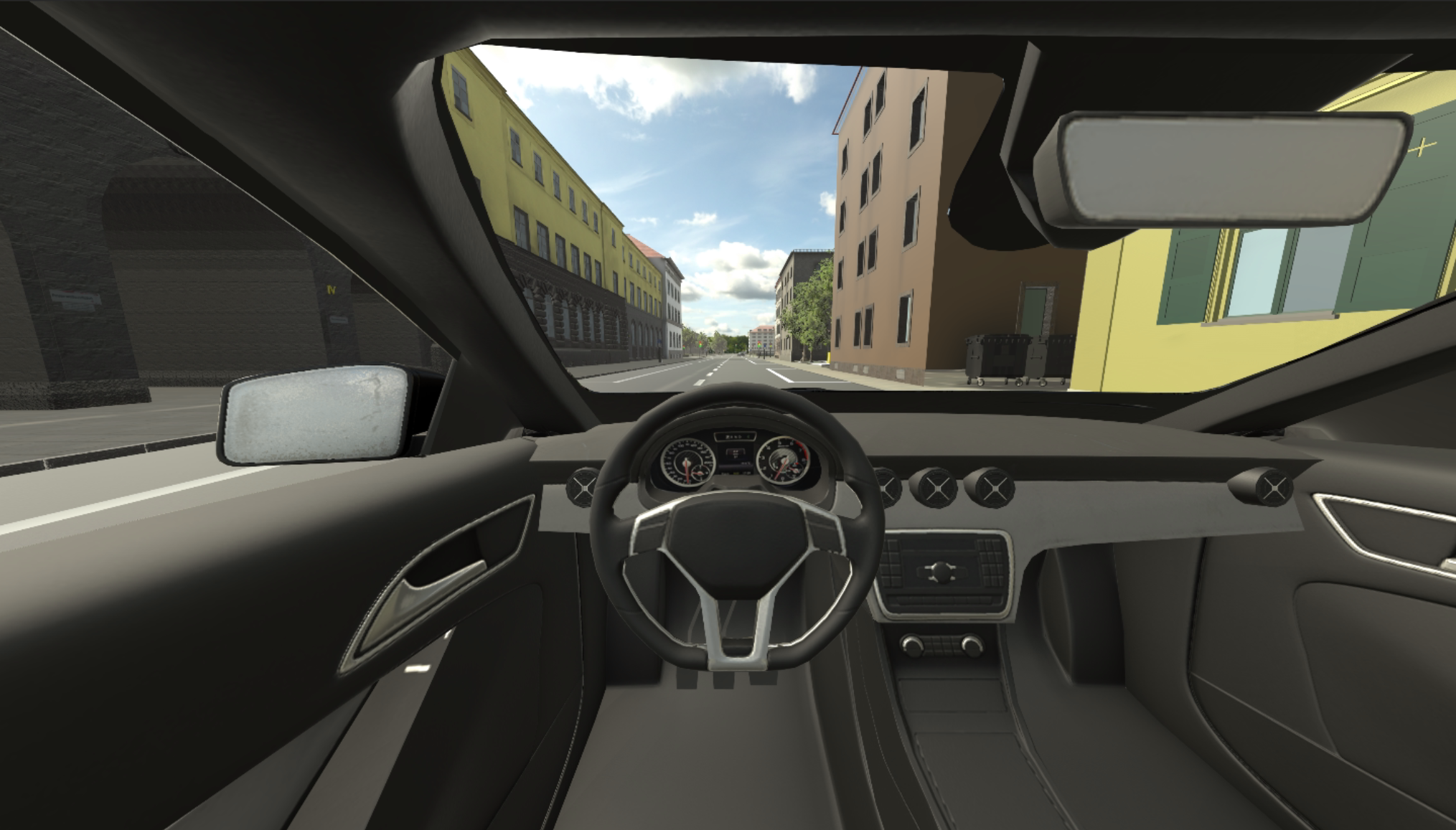}
    \caption{Runtime VR-Image rendered with Unity}
    \label{fig:VR_Car}
\end{figure}

In this subsection, we discuss the structure of the ADAS test procedure with external VR visualization using the Unity game engine.

To achieve a higher quality of simulation and visualization, the 3D city model of the Munich University of Applied Sciences is integrated into the conceptual model as a simulation environment. For this purpose, it will be imported into both software products. Since the original 3D model of Maxvorstadt has already been created as a project in Unity, the 3D model must consequently be exported from Unity, prepared, and imported into the simulation software Prescan. The Unity project represents the basis for the Unity VR visualization. It is essential that the positioning of the 3D model is identical in both software products to enable synchronization via coordinates.

The main framework components are the simulation software Prescan, the game engine Unity, the ROS server, and the real car (or the vehicle dynamic simulator). These components are planned to run on separate hardware to split the performance load. Prescan and Unity are each installed on a Windows computer. Since established ROS versions are only available for Linux systems, it is installed on such a computer and serve as the communication interface. Prescan natively supports a bidirectional connection to ROS systems, which allows data transfer from Prescan to the ROS server. This data transfer takes the form of a close to real-time data stream of all dynamic objects within the Prescan simulation. The future ROS2 release may improve even furthur communication performance.  

To connect the ROS server to Unity, an open-source Unity extension called ROS-Sharp is used, which has been developed by Siemens for robotics applications. With the connection established, vehicle data can thus be transferred from Prescan to Unity, whereby the conceptual model (see Fig. \ref{fig:DetailedToolchain}) technically enables bidirectional communication. Thus, vehicle data can also be sent from Unity to the simulation and influence it in real-time.

\begin{figure*} 
    \centering
    \includegraphics[width=1.0\textwidth]{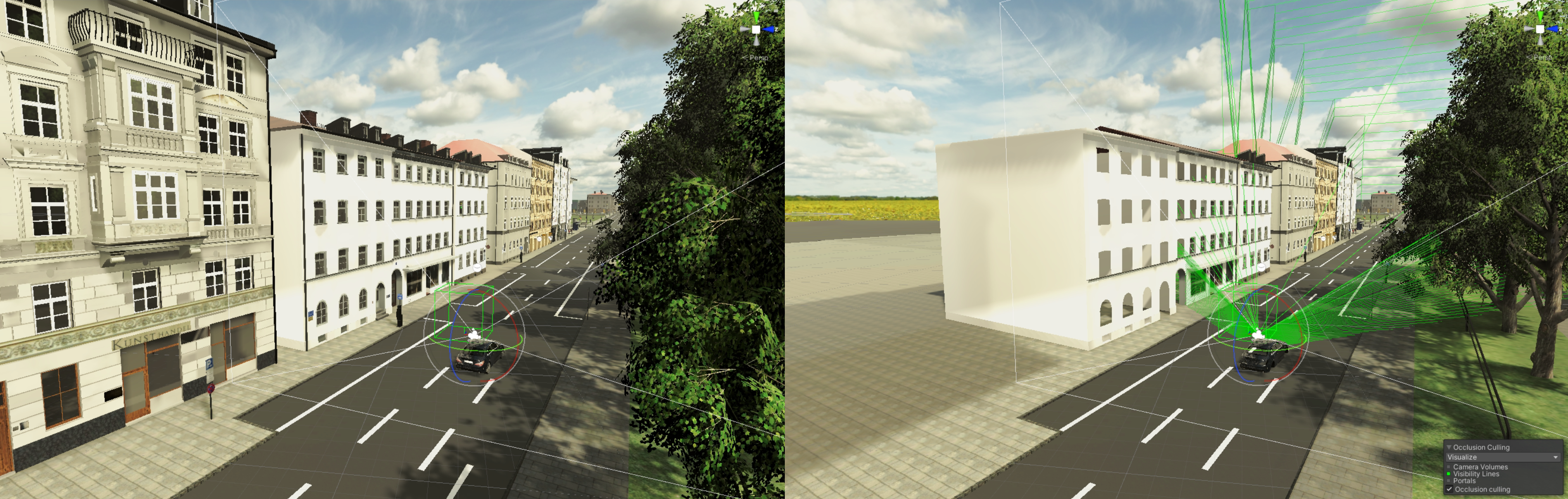}
    \caption{The effect of occlusion culling}
    \label{fig:OcclusionCulling}
\end{figure*}

\section{Optimization of the 3D Munich model for VR usage}
In this section, we will discuss the performance requirements for the implementation of VR visualizations.
The frame rate for smooth virtual reality visualizations should be around 90 frames per second (fps) and the maximum total cycle rate of the system must be less than 100 ms, within which the combined algorithm test can be considered real-time capable. Since special focus is placed on maintaining the immersion of the test driver, 100 ms is to be regarded as the limit value here \cite{Henning}.

 In addition, any improvements in the visual quality in Unity are generally associated with performance losses, which is why a basic performance of 100 to 200 fps is desirable for visually appealing VR visualizations. Especially with the focus on comfort testing, the quality of the visualization is essential to create immersion in the subject and thus obtain realistic reactions to different traffic driving scenarios. Virtual reality goggles manufacturers state a performance target of 72fps, 150-175 drawcalls per frame, and 300,000 to 1,000,000 polygons per frame \cite{OCULUS21}. VR goggles with higher refresh rates on the other hand also increase the performance requirements \cite{HTC21}.
Comparing these values with the performance values of the 3D model, it is noticeable that the current values deviate strongly from the target performance. Especially the polygon count of over 53,000,000 stands out and must be reduced. But also the number of drawcalls per image with over 340,000 are significantly above the target value for VR visualizations. The values refer to the polygon and drawcall counts are needed to render a representative section of the Munich model. A drawcall is created by executing a function on the graphics processing unit (GPU) to draw the screen at each frame. Modern 3D games with complex graphical assets and effects may require thousands of drawcalls to render the screen per frame \cite{LEHTOLA18}.
Thus, as the number of drawcalls increases, the workload of the GPU increases simultaneously, which in turn decreases the frames per second count of the visualization.

There are several approaches to reduce the number of drawcalls required to render a frame. The simplest approach is to remove drawn objects. This can be done by removing objects from the Unity scene - although this is not always possible, or by using the culling systems provided by Unity such as occlusion culling and frustum culling, or removing an LOD system \cite{LEHTOLA18}.

\subsection{Removing the LOD2 files from the city model}
The LOD2 models of the LDBV are performance-demanding as these are composed of countless single FBX files. It is desirable to isolate the 3D city model data from the interactive elements and other extensions, both to save performance and to make it easier to prepare the data for the simulation software Prescan. 

We could isolate 3D data from a Unity project, either by saving the 3D data as a unitypackage or exporting the data as an FBX file. 
Since the 3D data should only be separated from all unnecessary elements in the first step, saving it in a unitypackage is the best option.
Unitypackages are collections of files and data from projects or project elements that are compressed and saved in a file with the extension ".unitypackage". Similar to a zip file, an asset package retains its original directory structure and asset metadata when unpacked, making it a good way to share data between Unity projects or versions \cite{UNITY21}.
The package is consequently imported into a new Unity project of a more recent Unity version (2020.3.6f1 LTS). Note that the so-called LTS versions offer long-term support and thus ensure that new Unity versions will continue to be compatible with the project in the future.

\subsection{Reducing the number of polygons}
There are several techniques available to reduce the number of polygons in a Unity scene. The most obvious one is to reduce the polygons at the modeling level or to remove unnecessary houses in the rendering process of the 3D model. Since reducing the polygons of the individual models or houses using modeling software would require a great amount of manual effort, this method appears to be of little use. In addition, the dimensions of the virtual test scene are limited by the real test site anyway. For this reason, it is appropriate to reduce the 3D model to a specific, freely chosen section within the model, which leads to a significant reduction of polygons from 53.1 to 7.8 million.
This downsizing also results in a significant increase in frame rate from 2 to 40 fps. No 3D data has been deleted during the downsizing, but merely removed or hidden from the rendering process in Unity. This makes it flexible to change the selected sections, or to turn individual houses on or off even during runtime. 

\subsection{Reducing drawcalls}
The 3D model contains many different scripts for various purposes (first-person controller, sun position, menus, ...). Most of these scripts are not useful for the purpose of AD and ADAS algorithm testing and can be removed to reduce drawcalls.

 \subsubsection{Batching} In Unity, 3D objects (also called game objects) can be declared as static in the editor. This marking tells the game engine that the object will never move, and allows the game engine to render it with a decimated drawcall count using a method called batching. Batching means that multiple objects that have been assigned the same material are combined into a single draw call and drawn. Consequently, to take advantage of the performance benefits provided by this functionality, all non-moving elements are marked as static \cite{LEHTOLA18}.

   \subsubsection{ Occlusion culling} Besides to the default Frustum Culling option, which hides objects outside the camera view, there is an additional method of occlusion culling in Unity to further reduce the drawcall count. In fact, occlusion culling reduces the number of drawcalls within the camera view by removing objects that are hidden behind other objects from the rendering process \cite{INTEL14}. Unlike frustum culling, which is an automated process, occlusion culling must be configured and baked in the editor beforehand \cite{LEHTOLA18}.
The scene is divided into occlusion areas during so-called baking. The occlusion areas must be set up so that the camera view or test vehicle can never get outside of them. If the camera gets outside the occlusion areas, no occlusion culling can be applied, resulting in errors such as invisible objects and holes in the terrain. The accuracy of the culling can be adjusted by the size and frequency of the occlusion areas. Multiple small areas can be more accurate compared to a single large area but can increase the baking time and the size of the baked occlusion data file \cite{LEHTOLA18}. Fig. \ref{fig:OcclusionCulling} presents an effective example of occlusion culling. 

\begin{table}[b]
\small 
\captionsetup{size=small}
\setlength{\tabcolsep}{2pt} 
\begin{tabularx}{\columnwidth}{@{} l *{3}{C} c @{}} 
    \toprule
     Munich model & Original & Reduced size & Optimized performance\\
    \midrule
 	Number of polygons [million] & 53.1 &7.8& 1.3\\
 	Number of drawcalls & 342324  & 16676 & 4335\\
 	Frames per second [fps] &1.8 & 40.2 & 116.5\\
    \bottomrule
\end{tabularx}
\caption{Comparison of VR-optimized model performance}
\label{table:new}
\end{table}

   \subsubsection{LOD system} LOD is a system that can be used to improve the performance of a complex Unity scene. For this purpose, depending on the distance of the camera to the 3D objects, detailed 3D models (LOD3) are replaced by simplified 3D models (LOD2) or removed completely. In order not to affect the visual detail of the scene, this process should only take effect for 3D models that are far away. The LOD system is not limited to two LOD models and can be extended at will. However, this method for reducing draw calls is only applicable if corresponding LOD models have been modeled for the different 3D models and are available. Our Munich model includes LOD models for most of the buildings. 

Ultimately, Table I summarizes the presented optimization processes for VR usage. Overall, we have successfully reduces significantly the number of polygons and drawcalls, while at the same time improve frames per second and remain high quality visualization during VR testing for enhancing human perception.

\section{Conclusion}
We have presented a novel ADAS/AD testing and validation framework that attempts to improve human perception for perceived safety and comfort evaluation, and bridge the gap of two domains: ADAS engineering algorithms development and visualization. The technologies are realized via the development of an integrated modular VR visualization platform. Significant efforts have been also devoted to building, tuning, and optimizing the high quality 3D Munich city model. The model was integrated and optimized to both the ADAS simulation pipeline and the real-time VR platform, demonstrating the efficiency of the framework. We also discussed techniques to improve realism and performance of the VR visualization, which leads to an increased immersion level of the subjects and hence more meaningful evaluation of comfort. 

Future works will focus on furthur validating with standard and extensive ADAS testing scenarios, in particular traffic and safety-critical scenarios. The results will potentially help to reduce testing cost and time in the development stages. Moreover, motion sickness and other side factors when using VR glasses will be also be studied.

\section*{Acknowledgement}
This work is part of FOCETA project that has received funding from the European Union’s Horizon 2020 research and innovation programme under grant agreement No 956123.

\end{document}